\documentclass[aps,prd,reprint,groupedaddress,nofootinbib]{revtex4-2}

\makeatletter
\let\@fnsymbol\@arabic
\makeatother

\usepackage{graphicx,amssymb,amsmath,amsthm,amsfonts,epsfig,epsf,array,subfigure,appendix}
\usepackage[usenames]{color}
\usepackage[unicode=true,pdfusetitle, bookmarks=true,bookmarksnumbered=false,bookmarksopen=false, breaklinks=true,
colorlinks=true,citecolor=blue]
{hyperref}

\def\be{\begin{equation}}
\def\ee{\end{equation}}
\def\bea{\begin{eqnarray}}
\def\eea{\end{eqnarray}}
\def\bfla{\begin{flalign}}
\def\efla{\end{flalign}}

\def\gsim{\, \rlap{$>$}{\lower 1.1ex\hbox{$\sim$}}\,}
\def\lsim{\, \rlap{$<$}{\lower 1.1ex\hbox{$\sim$}}\,}

\definecolor{purple}{rgb}{0.7,0,1}

\definecolor{green}{rgb}{0,0.7,0.2}

\begin{document}

\title{
Black hole-wormhole collisions and the emergence of islands
}

\author{Jo\~{a}o M. Dias}
\email[]{joaotiagodias@tecnico.ulisboa.pt}
\affiliation{Departamento de F\'isica, Instituto Superior T\'ecnico -- IST, \\
Universidade de Lisboa - UL,
Avenida Rovisco Pais 1, 1049-001 Lisboa, Portugal}

\author{Antonia~M.~Frassino}
\email[]{antoniam.frassino@icc.ub.edu }
\affiliation{Departament de F{\'\i}sica Qu\`antica i Astrof\'{\i}sica, Institut de Ci\`encies del Cosmos,\\ Universitat de Barcelona, Mart\'{\i} i Franqu\`es 1, E-08028 Barcelona, Spain}

\author{Valentin~D.~Paccoia}
\email[]{valentindaniel.paccoia@studenti.unipg.it}
\affiliation{Dipartimento di Fisica e Geologia, Università degli Studi di Perugia, Via A. Pascoli, 06123, Perugia, Italy}

\author{Jorge~V.~Rocha}
\email[]{jorge.miguel.rocha@iscte-iul.pt}
\affiliation{Departamento de Matem\'atica, ISCTE -- Instituto Universit\'ario de Lisboa, Avenida das For\c{c}as Armadas, 1649-026 Lisboa, Portugal}
\affiliation{Centro de Astrof\'{\i}sica e Gravita\c{c}\~ao -- CENTRA, Instituto Superior T\'ecnico -- IST, \\
Universidade de Lisboa - UL,
Avenida Rovisco Pais 1, 1049-001 Lisboa, Portugal}

\date{\today}

\begin{abstract}
We use ray-tracing techniques to determine the evolution of the event horizon of a large black hole that ``gobbles'' a tiny, traversable wormhole. This calculation has physical meaning in the extreme mass ratio limit. Two setups are considered: a single-mouth wormhole connecting two otherwise independent universes, and a double-mouth zero-length wormhole within the same universe. In the first setting it turns out that, at early times, there exist two disconnected horizons, one in each universe, which then merge as the wormhole falls into the large black hole. In the second setup, we observe the appearance of an `island', a region of spacetime that is spatially disconnected from the exterior of the black hole, but in causal contact with future null infinity. The island shrinks as time evolves and eventually disappears after sufficient time has elapsed, as compared to the distance between the two mouths. This provides a communication channel with the interior of the large black hole for a certain time interval. We compute numerically the lifetime of the island and verify that it depends linearly on the inter-mouth distance. Extending the analysis to wormholes with finite length, we show that the achronal averaged null energy condition prevents the appearance of islands.
\end{abstract}



\maketitle

\section{Introduction
\label{sec:Intro}}

Over the past few years, the LIGO/VIRGO collaboration has provided a wealth of information about the dynamics and properties of highly compact objects such as black holes (BH) and neutron stars~\cite{LIGOScientific:2021djp}. These observations have confirmed many predictions of Einstein's theory of general relativity (GR) and sparked a strong interest in better understanding the merging of compact objects. 
Other experimental observations regard the shadow of black holes and indirect measures of them instead~\cite{EventHorizonTelescope:2019dse}. However, none of these measurements offers any information about what happens in the interior of a black hole.  

Once GR is coupled to matter, additional solutions arise. Notably, topologically non-trivial spacetimes, such as wormholes (WH), can be found~\cite{Ellis:1973yv, Bronnikov:1973fh, Morris:1988cz, Morris:1988tu, Visser:1995cc}. These are structures that connect two distant regions of space-time, providing a possible shortcut for travel through the universe, or even two otherwise disconnected universes. 
However, the existence of wormholes typically requires exotic or negative-energy matter, violating the standard energy conditions~\cite{Morris:1988cz, Agnese:1995kd, Visser:1997yn, Visser:2003yf, Lobo:2007zb, DeFalco:2021ksd}. This has been a significant obstacle to the viability of wormholes. Nevertheless, several ways to overcome this issue have been discussed over the past years, including classical solutions of general relativity coupled to massless charged fermions~\cite{Maldacena:2018gjk, Fu:2019vco} or massless scalar fields~\cite{Barcelo:1999hq, Agnese:1995kd}, alternative (higher derivative) theories of gravity~\cite{Hochberg:1991tz, Hochberg:1990is, Lobo:2009ip},  models involving extra dimensions~\cite{Kar:2015lma,Bronnikov:2002rn,Maldacena:2020sxe} or backreaction from quantum fields~\cite{Hochberg:1996ee} and quantum gravity~\cite{Kundu:2021nwp}. These are only some of the possibilities that are currently being explored to understand better the nature of wormholes and their potential theoretical implications.

In this paper, we are interested in studying the shape and features of the event horizon when a wormhole plunges into a black hole.
To the best of our knowledge, the idea of using a wormhole as a means to access the interior of a black hole was first pointed out and investigated by Frolov and Novikov~\cite{Frolov:1993jq}.
Most of the details worked out in that early study relied on a simplified model to describe the infalling wormhole, in the sense that it was taken to be point-like, while also considering a quasi-Newtonian approximation to describe the gravitational field in the vicinity of the wormhole mouths.
Here we will obtain the full fledged evolution of the event horizon as it responds to the presence of an extended wormhole. In the process, we clarify certain results of the Frolov-Novikov model, and point out novel phenomena occurring in this setting: the possibility of a black hole region \emph{exterior} to an event horizon, and the appearance of an ephemeral island. We shall see that the former feature is present when the wormhole connects to a different universe, while the latter feature arises when the wormhole exhibits two mouths within the same universe.  

The analogous problem of the fusion of two black holes has been studied in a regime in which the ratio of their masses, $m$ and $M$, is very small ---the extreme mass ratio limit (EMR). The full evolution of the event horizon in such a setting can be obtained exactly and with elementary techniques when $m/M \rightarrow 0$~\cite{Emparan:2016ylg}.
The key idea is that, in the rest frame of the large black hole, and at scales much shorter than its mass (in fact, taking $M \rightarrow \infty$), the small black hole is freely falling in flat space while the large black hole horizon can be thought of as just an acceleration horizon.

Returning to the case of a BH-WH merger, the extreme mass ratio limit was also implicitly considered in~\cite{Frolov:1993jq} but in that study it was the mass of the large black hole $M$ that was kept fixed, while the size of the wormhole was taken to zero. In that context, one has to consider ---and speculate about--- what happens when the wormhole finally hits the black hole singularity. When taking the extreme mass ratio limit as we shall do, the BH singularity is infinitely far away and the wormhole never really reaches it. Instead, we will be able to capture the time evolution of the horizon as it falls \textit{through} the wormhole throat. This is only possible if the size of the wormhole is kept finite.

Therefore, this approximation reveals the spacetime geometry of the merger and the determination of the event horizon reduces to finding the congruence of null geodesics that approaches a null plane at late times~\cite{Emparan:2016ylg}.
This limit provides a boundary condition on the null generators of the event horizon and allows one to integrate the geodesic equations back in time up to the caustic points (the set of points where the horizon generators focus). This same approach was also employed to study mergers with rotating~\cite{Emparan:2017vyp} or charged~\cite{Pina:2022dye} black holes, as well as the fusion of a neutron star with 
a large black hole~\cite{Emparan:2020uvt}.

Hereafter, we apply these well-established techniques to the study of the merger of a small wormhole with a large black hole in the EMR limit. 
We will adopt the Ellis-Bronnikov~\cite{Bronnikov:1973fh, Ellis:1973yv} spacetime to model the wormhole. 
As mentioned, wormholes can either connect two different universes or two distant regions in the same universe.
A single solution of Einstein's equations can be used to tackle both types of wormhole described, the essential difference being that they differ at the level of the topology of the global spacetime.
In this paper, we compute the null generators considering both these topologies in order to determine the evolution of the associated event horizon.  
In particular, in Sec.~\ref{sec:ebwormhole}, we introduce the specific wormhole metric we will use for our calculations. The method adopted is explained in detail in Sec.~\ref{sec:BH_merger_description} while in Sec.~\ref{sec:inter_universe} and Sec.~\ref{sec:intra_universe} the results for the merger of a large BH with inter- and intra-universe wormholes, respectively, are presented. Finally, we conclude and discuss future prospects in Sec.~\ref{sec:conclusions}.
Throughout this work, we adopt geometrized units, for which $c=1$.

\section{Ellis-Bronnikov wormhole \label{sec:ebwormhole}}

The static spherically symmetric Ellis-Bronnikov (EB) metric is a simple, special case of a traversable wormhole solution found in 1973~\cite{Bronnikov:1973fh, Ellis:1973yv} and reads
\begin{equation}
    ds^2 = - dt^2 + dR^2 + (R^2 + a^2) (d \theta^2 + \sin^2 \theta d\phi^2)\,,
\end{equation}
where the parameter $a$ determines the size of the throat and the radial coordinate ranges in $R \in (-\infty, +\infty)$. Note, however, that $R$ does not correspond to an areal radius. The throat occurs at $R=0$, and the two limits $R\rightarrow \pm \infty$ correspond to the two asymptotically flat regions connected through the wormhole throat. The metric has no horizons.
A change of coordinates, $r^2=R^2 + a^2$, shows that the resulting solution 
\begin{equation} \label{EllisBmetric}
    ds^2 = -dt^2 + \frac{r^2}{r^2 -a^2} dr^2 + r^2 (d\theta^2 + \sin^2 \theta d\phi^2)\,,
\end{equation}
is of the Morris-Thorne form~\cite{Morris:1988cz}
\begin{equation}
    ds^2 = - e^{2 \Phi(r)} dt^2 + \frac{dr^2}{ 1-b(r)/r} + r^2 (d \theta^2 +\sin^2 \theta d\phi^2)\,,
\end{equation}
with  $\Phi(r)=0$ and $b(r) = a^2/r$.
 The metric~\eqref{EllisBmetric}, corresponding to a wormhole with zero gravitational mass,
 has been studied in detail in \cite{Bugaev:2021dna} with the purpose of investigating the bending and scattering of light rays passing near the wormhole as well as ray capture and wormhole shadows. It has also been studied in~\cite{Ohgami:2015nra,Ohgami:2016iqm} to understand the shadows of the EB wormhole surrounded by nonrotating and rotating dust.

A comment is in order. Since the wormhole has zero mass, the ratio between the masses of the small wormhole and the large black hole is trivially zero. Therefore, the terminology ``extreme mass ratio'' cannot strictly be applied in this context. Nevertheless, one can still define a small parameter using dimensionful quantities associated to both the wormhole and the black hole. Instead of their masses, we consider corresponding length scales: for the black hole one takes the horizon radius, and for the wormhole we use the throat radius. However, we shall continue to refer to the regime of a binary consisting of a small wormhole and a large black hole as the extreme mass ratio limit, which is widespread nomenclature.
In practice, we shall take the formal limit of the BH horizon radius tending to infinity, while keeping the radius $a$ of the throat fixed.

\section{Description of the merger
\label{sec:BH_merger_description}}

As already mentioned, the merger will be completely determined by the small object's metric.
The starting point to identify the causal horizon in such a geometry is Hamilton's variational principle for extremal spacetime paths, that leads to the equations
\begin{equation}
    \frac{d^2x^\mu}{d\lambda^2}+\Gamma^\mu_{ \nu\rho}\frac{dx^\nu}{d\lambda}\frac{dx^\rho}{d\lambda}=0\,,
\end{equation}
which are none other than the geodesic equations, where $\lambda$ is an affine parameter. 
Hamilton's equations, in terms of the impact parameter $q=L/E$, for spherically symmetric metrics of the type
\begin{equation}
    \label{genericmetricss}
    ds^2
    =-f(r)dt^2+\frac{dr^2}{g(r)}+r^2\left(d\theta^2+\sin^2{\theta}d\phi^2\right)\,,
\end{equation}
reduce to \cite{Emparan:2020uvt}
\begin{gather}
\begin{aligned}\label{hameqsymm}
    \Dot{t}&=f(r)^{-1}\,,\\
    \Dot{\phi}&=-\frac{q}{r^2}\,,\\
    \Dot{r}&=g(r)p_r\,,\\
    \Dot{p_r}&=-\frac{f'(r)}{2f(r)^2}-\frac{g'(r)}{2}p_r^2+\frac{q^2}{r^3}\,,
\end{aligned}
\end{gather}
where $p_r$ is the conjugate momentum associated to the radial variable $r$, dots refer to derivatives with respect to the normalized affine parameter $E\lambda$, and primes represent derivatives with respect to $r$. 
The spherical symmetry of the setup allows one to choose the coordinate system in such a way that geodesics are always contained in an equatorial plane\footnote{However, note that the equatorial plane for distinct geodesics is, in general, different.}. 

At this point, the only missing step towards the numerical integration and, therefore, the determination of the event horizon of the EMR merger, is the identification of the asymptotic behavior of the solutions at large distances, which also correspond to large values of $\lambda$ (i.e., $\lambda \rightarrow \infty$).
It is convenient to use the radial coordinate $r$ to parametrize the geodesics, instead of $\lambda$. In that case, once functions $f(r)$ and $g(r)$ have been given, the asymptotic behavior can be extracted by expanding in powers of $1/r$, and integrating Eqs.~\eqref{hameqsymm}, thus yielding the coordinates $t$ and $\phi$ as functions of $r$. The integration constants are then fixed by demanding the property, already mentioned, that they should asymptote a null plane. We can then continue the integration of the geodesics numerically back in time.

In certain cases, it will happen that these null geodesics intersect. Such intersection points are spacetime events from which light rays emitted to the future in several directions belong to the event horizon. In other words, these are points where new generators enter the horizon. These points are referred to as {\it caustics}. Generally they show up as a continuous set, forming {\it caustic lines} when they are one-dimensional. Note that caustic lines are \emph{not} geodesics.  Given the high degree of symmetry involved in the setup, it is clear that there must exist a caustic line along $\phi=\pi$. When a geodesic hits that axis, at a given $r=r_{caustic}$, we terminate the numerical integration.

\begin{figure}[t!]
\centering
\includegraphics[width=0.95\linewidth]{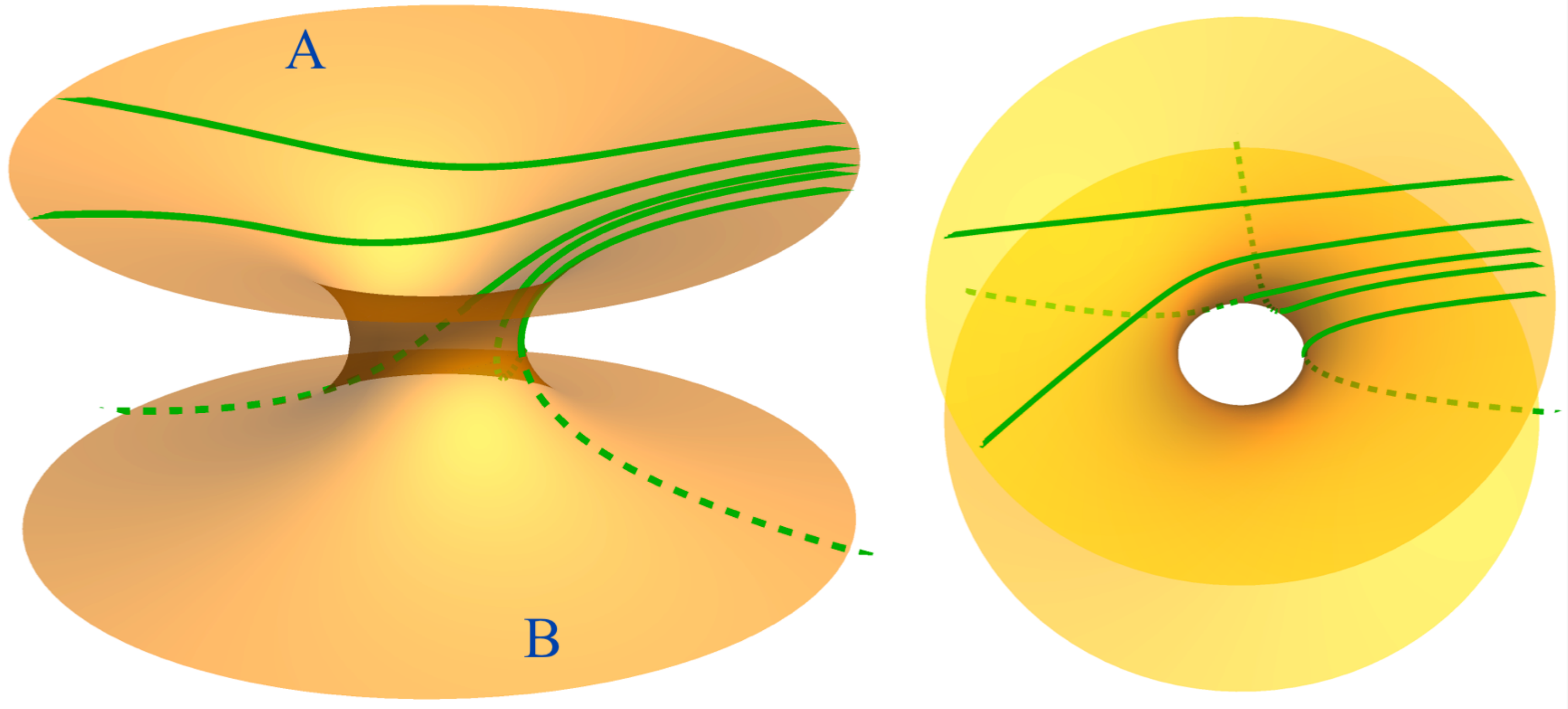}
\caption{Embedding diagram of the inter-universe wormhole, from two different perspectives. The wormhole connects the two universes (named A and B) through the throat. Two kinds of geodesics (green lines) that end up at infinity in universe A are shown in the figure: geodesics that always remain in the A side; geodesics that come in from infinity in the B side and emerge from the throat of the wormhole. The event horizon of the large black hole is represented by the (continuous) family of geodesics considered.}
\label{fig:interWH}
\end{figure}

Another subtlety one has to deal with is the existence of {\it turning points} for some of the geodesics. These arise as points of closest approach between a geodesic and the wormhole (see Fig.~\ref{fig:interWH}), and we shall denote their radial coordinate by $r_{min}$. At $r=r_{min}$, the derivative $d\phi/dr$ necessarily blows up, so the solution jumps to another branch. In practice, the continuation of the integration after the turning point is related, by symmetry arguments, with the solutions before the turning point ~\cite{Emparan:2016ylg},
\begin{align}
\label{turning_points_phi}
    \phi^{\rm (after)}(r)&=2\phi(r_{min})-\phi^{\rm (before)}(r)\,,\\
\label{turning_points_t}
    t^{\rm (after)}(r)&=2\,t(r_{min})-t^{\rm (before)}(r)\,,
\end{align}
with the understanding that these expressions only apply in the range $r\in[r_{min},r_{caustic}]$.

Although we perform our computations using spherical coordinates, it is useful to convert to Cartesian coordinates,
\begin{equation}
    \label{xz}
    x=r\sin{\phi}\,,\quad z=r\cos{\phi}\,,
\end{equation}
to visualize the results. The third Cartesian coordinate, $y$, is forced to vanish since each geodesic is constrained to lie on an equatorial plane.
To consider the time evolution of the event horizon, we draw the congruence of generators as a two-dimensional surface in the three-dimensional $(t,x,z)$ spacetime.

\begin{figure*}[t]
\centering
\includegraphics[width=0.85\linewidth]{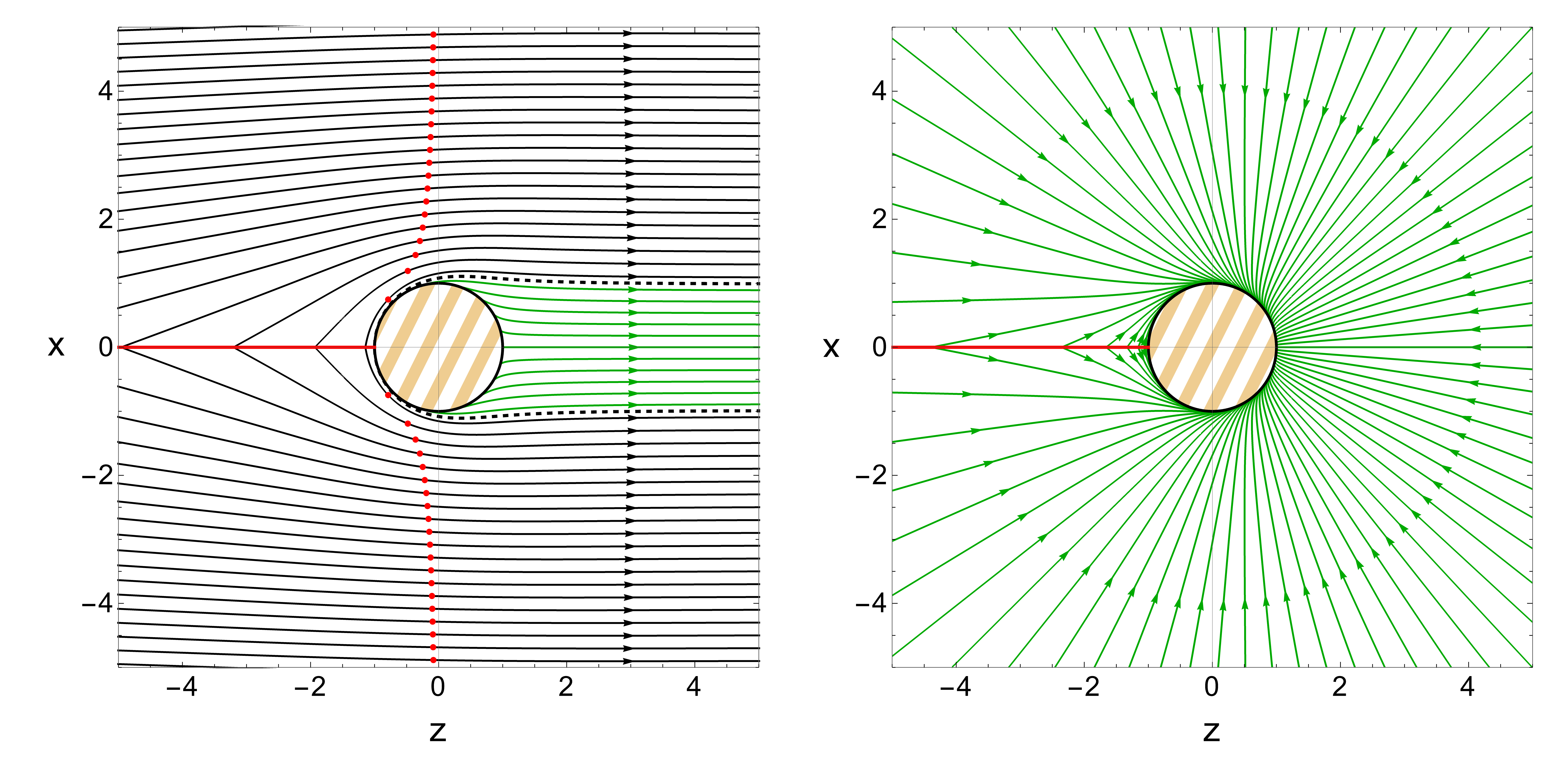}
\caption{Projection of the generators of the event horizon in the $(x,z)$ plane, as defined in Eq.~(\ref{xz}). The left (right) panel represents the null geodesics in universe A (universe B). The small arrows indicate the flow of time along each geodesic. All rays on side A propagate, to the future, towards $z\to+\infty$. Green curves are generators emerging from the wormhole and their specific behavior on side B is determined by imposing continuity and differentiability at the throat. The dotted black curves correspond to the threshold value of the impact parameter, $q=q_c$, which distinguishes the geodesics that cross the wormhole (in green) from those that remain always outside (in black). Black curves correspond to (caustic) generators with $q>q_c$. The red dots in the left panel indicate turning points, i.e., points along the generators at which the distance from the wormhole is minimized. In both panels, the red lines correspond to the caustic line.}
\label{fig:projection_bothsides}
\end{figure*}

\section{Merger of a black hole with a single-mouth wormhole\label{sec:inter_universe}}

In this section we will compute the generators that determine the event horizon in the collision of a large black hole and a single-mouth wormhole, in the extreme mass ratio limit.

The setup considered is shown in Fig.~\ref{fig:interWH}. It illustrates two universes (sides A and B) which are put in causal contact due to the existence of a wormhole between them. We assume that we live on side A and hence the boundary conditions mentioned in the previous section will be imposed on that side of the wormhole only. Integrating backwards in time, two qualitatively different geodesics can arise: those that always remain in universe A and those that traverse the wormhole into side B.

It is clear, from a simple continuity argument, that all geodesics crossing the throat have an impact parameter smaller than a certain threshold value, $q_c$. The $S^2$ throat is completely fibered by such geodesics which cross it at varying angles. The outcome is that on side B the generators come out at all possible angles, revealing a roughly radial structure that is quite different from the congruence observed on side A (see Fig.~\ref{fig:projection_bothsides}).

We now turn to the actual computation of these geodesics.

\subsection{Horizon generators
\label{subsec:inter_universe_generators}}

\begin{figure*}[t!]
\centering
\includegraphics[width=0.85\linewidth]{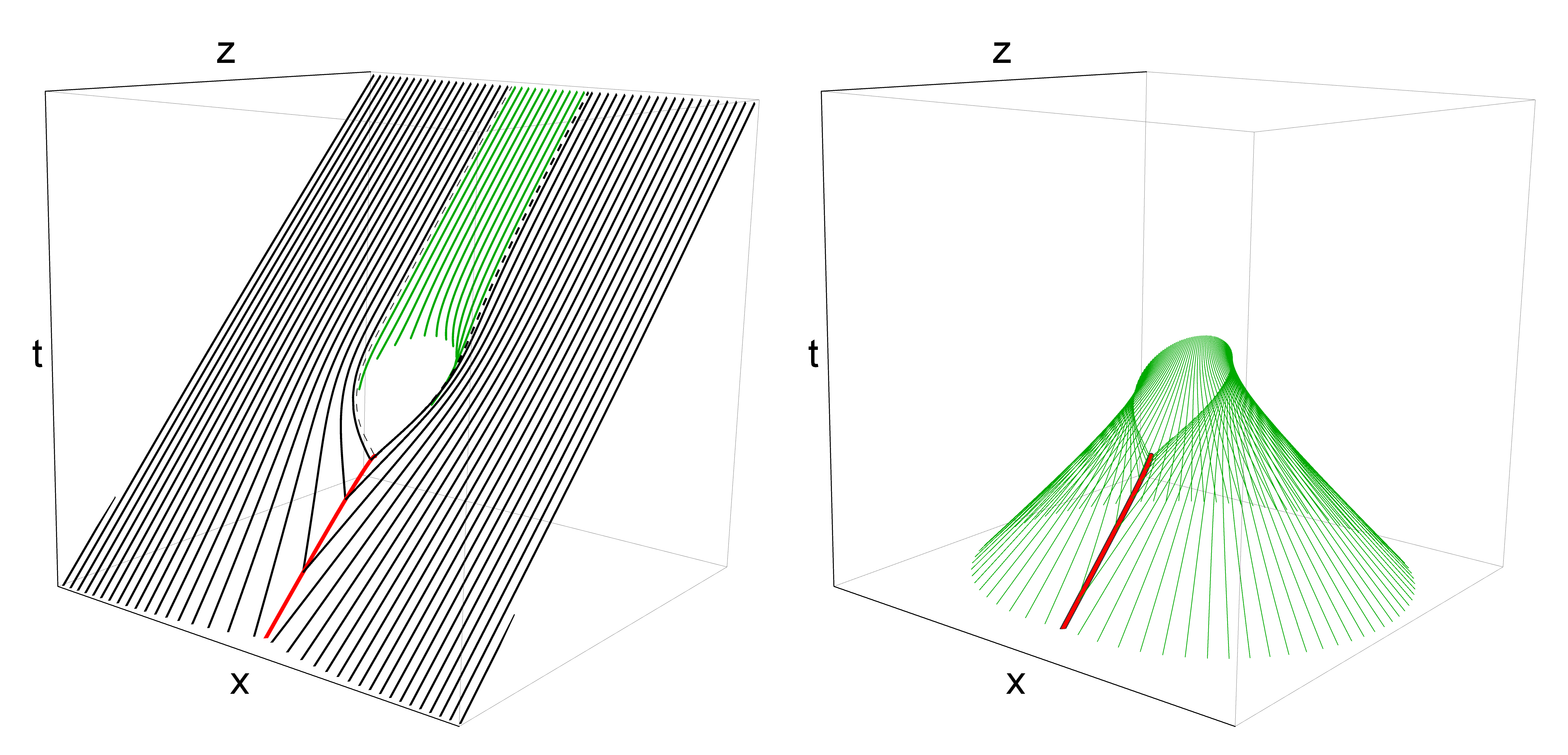}
\caption{Left panel: event horizon in the equatorial plane (i.e., the third spatial coordinate ---not shown--- is set to $y=0$) as a function of time. Green curves indicate generators that cross the wormhole. The dashed black curves correspond to the critical generators with impact parameter $q=q_c$. Black curves correspond to caustic generators with $q>q_c$. Right panel: causal horizon on side B of the universe (and in the equatorial plane) as a function of time. Events below the green cone can influence side A of the universe, where the merger is happening, outside of the large black hole's event horizon. Signals emitted from events to the future of the cone will either remain on side B or else, if they cross the wormhole, they will inevitably end up in the large black hole interior on side A. The red curve indicates the caustic line. It is apparent that the line of caustics actually continues through the wormhole into the other universe.}
\label{fig:ebmtmerger}
\end{figure*}

Using the metric~\eqref{EllisBmetric}, we see that the first two equations~\eqref{hameqsymm} for the light rays reduce to
\begin{equation}\label{conservedm2}
    \Dot{t}=1\,,\qquad \Dot{\phi}=-\frac{q}{r^2}\,.
\end{equation}
Instead of using the remaining two equations to determine the expression for $\Dot{r}$, it is simpler to obtain it from the quantity $\epsilon=-g_{\mu\nu}\Dot{x}^\mu\Dot{x}^\nu$, which is conserved and equal to $0$ along a null geodesic. Explicitly, this equation reads
\begin{equation}
\label{ldotWH}
    -\Dot{t}^2+\left( \frac{r^2}{r^2-a^2}\right) \Dot{r}^2+r^2\Dot{\phi}^2=0\,.
\end{equation}
Multiplying by $(r^2 - a^2)/r^2$ and using Eqs.~(\ref{conservedm2}) results in 
\begin{equation}\label{rdoteb}
    \Dot{r}=\frac{1}{r^2}\sqrt{\left( r^2-q^2\right) \left( r^2-a^2\right)}\,.
\end{equation}

Directly integrating this equation would give $\lambda(r)$; however, this results in a combination of elliptic integrals which cannot be inverted analytically to find $r(\lambda)$ and thus $\phi(\lambda)$. This problem is circumvented by using $r$ instead of $\lambda$ as an independent variable, by rewriting the integrals as
\begin{equation}\label{ints}
t_q(r)=\int dr\, \frac{\dot{t}}{\dot{r}}\,,\qquad \phi_q(r)=\int dr\,\frac{\dot{\phi}}{\dot{r}}\,,
\end{equation}
where the label $q$ indicates the (only) parameter of this family of geodesics. The requirement that this congruence has to approach a planar horizon at infinity is the boundary condition that fixes the integration constants in (\ref{ints}).

Now we expand Eqs.~\eqref{conservedm2} and~\eqref{rdoteb} at large distances, $r \gg a$ (corresponding to large values of $\lambda$). 
For $\phi_q$, we have
\begin{equation}
    \phi_q(r\to\infty)=\int dr\,\frac{\dot{\phi}}{\dot{r}}\biggr\rvert_{r\rightarrow \infty}=\alpha_q+\frac{q}{r}+\mathcal{O} \left( r^{-3} \right) \,.
\end{equation}
We fix the integration constant $\alpha_q=0$ in such a way that asymptotically the geodesics are aligned with the $z$ axis.
For $t_q$, we get
\begin{equation}
    t_q\left( r\to\infty\right) = r+\beta_q+\mathcal{O}\left( r^{-1}\right)\,.
\end{equation}
Requiring that all generators approach a null plane demands that $\beta_q$ is a constant independent of $q$. Since the line element~\eqref{EllisBmetric} is static, we can fix $\beta_q=0$ without loss of generality.

The general results for Eqs.~\eqref{ints} are given in terms of elliptic integrals\footnote{It is important, when evaluating these expressions, to be careful with the prescription for the square root of complex numbers and with the branch cuts in the elliptic functions. The prescriptions used here are those implemented in \textit{Mathematica 12}.},
\begin{align}
    t_q(r)&=\int\frac{r^2dr}{\sqrt{(r^2-q^2)(r^2-a^2)}}\,,
\label{eq:integ_t}\\
    \phi_q(r)&=-\int\frac{q\,dr}{\sqrt{(r^2-q^2)(r^2-a^2)}}\,,
\label{eq:integ_phi}
\end{align}
but, in practice, we integrate the expressions for $dt_q/dr$ and $d\phi_q/dr$ numerically, starting from a very large radius.
It is possible to obtain analytic expressions but they are cumbersome, and therefore we relegate them to the Appendix~\ref{app:AppendixA}. We employ them to compute explicitly the caustic line. Nevertheless, note that the central generator, having $q=0$, has a very simple form:
\begin{align} \label{eq:centralg}
    t_{q=0}(r)&=\sqrt{r^2-a^2}\,,\\
    \phi_{q=0}(r)&=0\,.
\end{align}

Because the generators with $q<q_c$ fall through the wormhole, it is important ---in order to have the complete picture of the problem--- to follow them back in time on the \emph{other side} of the wormhole. This is done by spatially reflecting the solutions already found in the A universe\footnote{This procedure is extremely similar to the continuation~(\ref{turning_points_phi}-\ref{turning_points_t}) of the geodesics past the turning points.}. Alternatively, for each geodesic one can perform a new set of integrations on the B universe, using as initial conditions the endpoints of the previous integrations in the A side. Continuity of the geodesic across the throat is easily achieved by starting the integration from the same point on the throat. Continuity of the {\it velocity} also demands that the crossing angle be the same on both sides of the throat. 
More precisely, a ray entering the wormhole at a certain angle relative to the tangent of the wormhole, will exit on the other side at the supplementary angle compared to the tangent at that same point:
\begin{align}
    \phi_{q}^{(B)}(r)&=2\phi_q(a)-\phi_q(r)\,,\\
    t_{q}^{(B)}(r)&=2t_q(a)-t_q(r)\,,
\end{align}
where the superscript $B$ refers to the coordinates in the B universe.

\subsection{Physical interpretation}

In the alternative universe, the null generators continue into the past as an expanding cone, and the line of caustics also extends along that cone. This can be seen in Fig.~\ref{fig:ebmtmerger}. This emerging cone can be regarded as a \emph{causal horizon}, although in this context its interpretation is quite unconventional.
Events \emph{inside the cone} are those which can influence universe A in the region outside the large black hole, i.e., light rays emitted from points inside this cone can escape to future null infinity (${\cal I}^+$) on side A.
On the other hand, events \emph{outside the cone} can only influence the A universe in the region inside the large black hole.
It should be stressed that light rays originated in the B side ---whether inside or outside the cone--- can always escape to ${\cal I}^+$ on side B.
Similarly, events on side A ---whether inside or outside the black hole--- are also in causal contact with ${\cal I}^+$ on side B.

\begin{figure}[t!]
\centering
\includegraphics[width=0.95\linewidth]{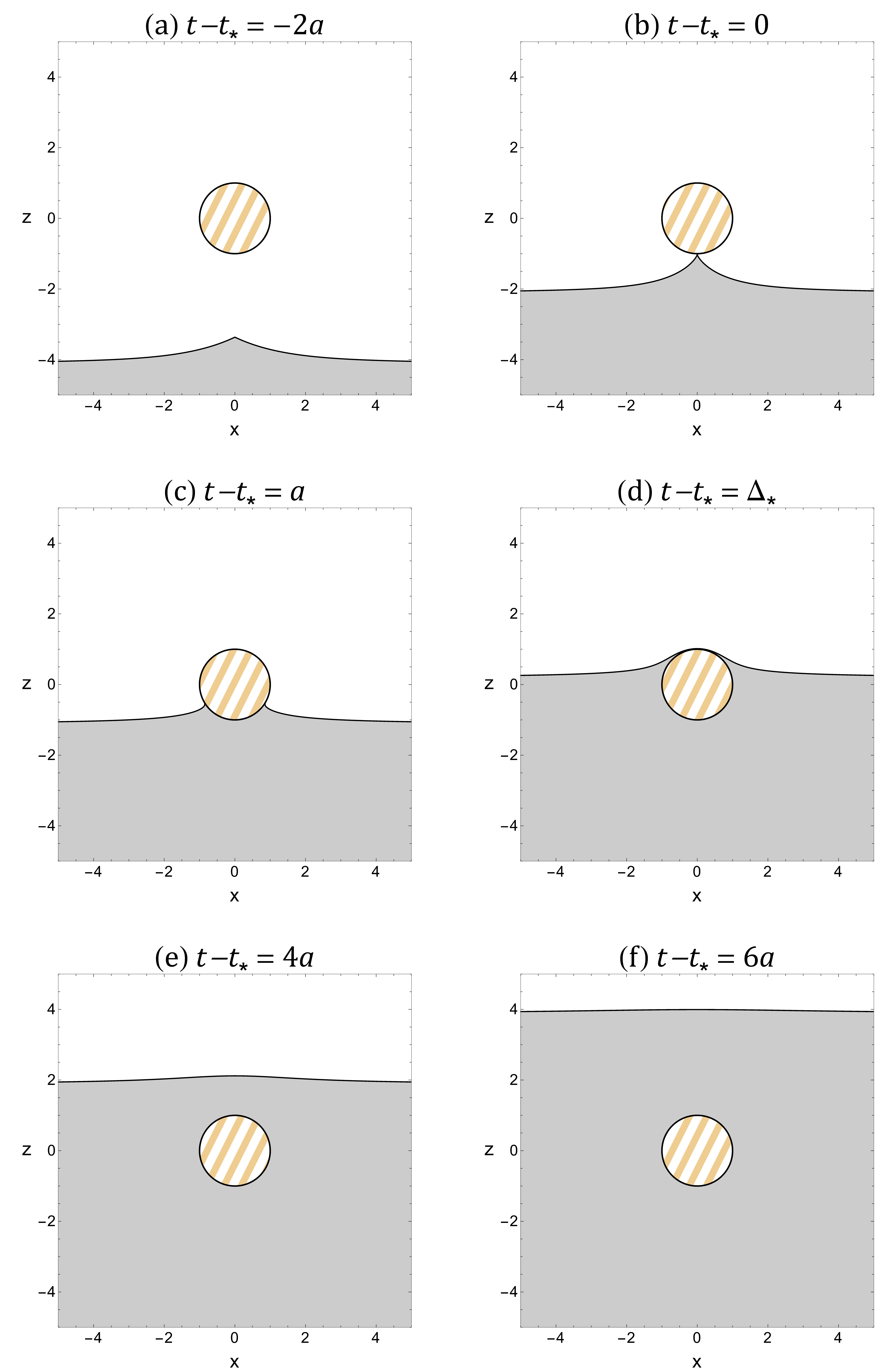}
\caption{Sequence of constant-time slices of the event horizon in the A universe (from the left panel of Fig.~\ref{fig:ebmtmerger}) with spatial coordinates centered on the small Ellis-Bronnikov wormhole. The shaded gray area represents the interior of the black hole. Pinch-on occurs at $t=t_*$ $(b)$. The time interval $\Delta_*\simeq2.31244\,a$ in panel $(d)$ is a natural measure of the duration of the fusion. The full two-dimensional constant-time slices of the event horizon are obtained by rotating around $x=0$. Axes are in units of $a=1$.}
\label{fig:ebslices}
\end{figure}

\begin{figure}[t!]
\centering
\includegraphics[width=0.95\linewidth]{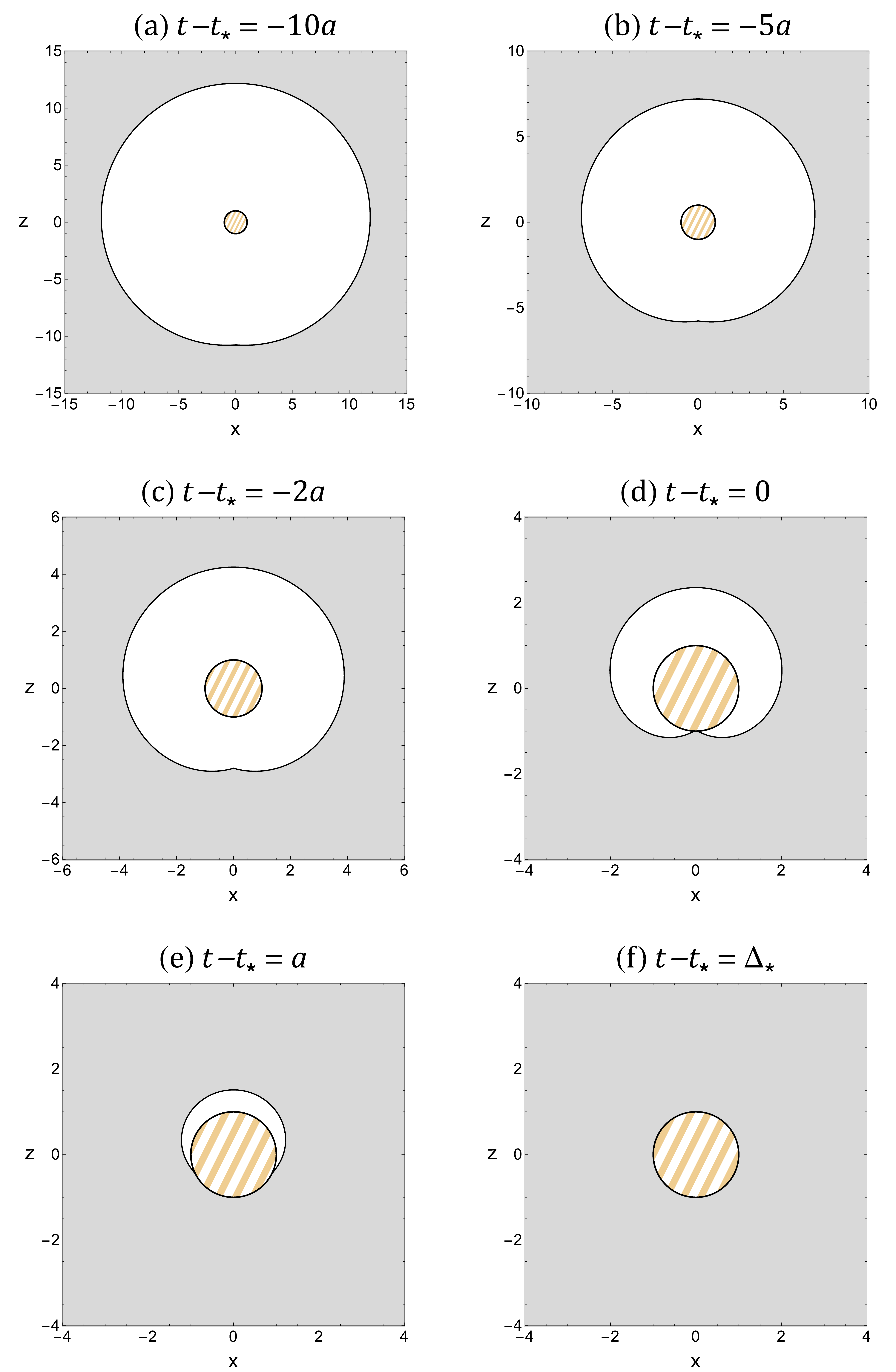}
\caption{Sequence of constant-time slices of the event horizon in the B universe (from the right panel of Fig.~\ref{fig:ebmtmerger}), with spatial coordinates centered on the small Ellis-Bronnikov wormhole. The shaded gray area represents the interior of the black hole. Pinch-on occurs at $t=t_*$ $(d)$, at which time the kink along the caustic has already fell through the wormhole (pinch-on occurs very close to the wormhole throat but slightly shifted to side A of the universe). After a time interval $\Delta_*=2.31244\,a$ the entire universe B is inside the black hole (panel $(f)$). Axes are in units of $a=1$.}
\label{fig:slicesB}
\end{figure}

The resulting picture is that in a BH-WH collision, as in a binary black hole merger, there are two initially disconnected event horizons that fuse (at the `pinch-on' time) to become a single connected surface at late times. The crucial difference is that the `small' horizon lives almost entirely in the B universe, and as we go back in time this topologically spherical horizon expands at the speed of light. In fact, from the perspective of an observer living in universe A, the (large) black hole corresponds to the grey region in the bottom part of Fig.~\ref{fig:ebslices} {\em together with} the grey region in the {\em exterior} of the event horizon traced in universe B (see Fig.~\ref{fig:slicesB}).
This is a peculiarity that stems from the fact that the spacetime considered possesses two asymptotic flat ends (one in side A, and another in side B, as apparent in Fig.~\ref{fig:interWH}).

\subsection{Quantitative characterization}
\label{subsec:QC_inter}

It is possible to distinguish between rays that enter the horizon at a caustic with $q>q_c$ (black geodesics in Fig.~\ref{fig:projection_bothsides}) and rays that emerge from the wormhole with $q<q_c$ (green geodesics). The critical value $q=q_c$ corresponds to rays that go through the point $(r,\phi)=(a,\pi)$, which implies the equation
\begin{equation}
    \phi_{q_c}(a)=\pi\,.
\end{equation} 
This can be solved numerically to find 
\begin{equation}
    q_c=0.985821\ a\,.
\end{equation}

In previous calculations~\cite{Emparan:2016ylg}, a parameter $q_{*}$ has been introduced to distinguish between generators which enter the horizon on the side of the large black hole, $q>q_*$, and generators with $q_c<q<q_*$ which enter on the side of the small object. 
The value of $q_*$ (and its associated value $r_*$) is determined by the condition $\dot{r}\rvert_{\phi=\pi}=0$, together with Eq.~(\ref{rdoteb}), which yields
\begin{equation}
\label{eq:rdotequalzero}
    \left(r_*^2-q_*^2\right)\left(r_*^2-a^2\right)=0\,,\qquad \phi_{q_*}(r_*)=\pi\,.
\end{equation}
These equations have two acceptable solutions. One of them is trivially $q_*=q_c\,,\; r_*=a$, corresponding to the critical impact parameter that reaches the caustic line $\phi=\pi$ exactly at the wormhole throat.
Using the expression~\eqref{eq:nuvsxi} from the Appendix~\ref{app:AppendixA}, one can compute numerically the other solution to be
\begin{equation}
\label{eq:qstar2}
    q_*=r_*=1.01581\, a\,.
\end{equation}
This second solution is, in fact, the relevant one that separates the generators that enter the horizon on the side of the large black hole from those that enter on the side of the small wormhole. This is more easily understood by inspecting the figures in the Appendix, which are zoomed-in versions of Figs.~\ref{fig:projection_bothsides} and~\ref{fig:ebmtmerger}.
The corresponding value of time, $t_*$ (see panel $(b)$ in Fig.~\ref{fig:ebslices}, or panel $(d)$ in Fig.~\ref{fig:slicesB}), can be obtained by inserting these values into Eq.~\eqref{eq:etavsnu}:
\begin{equation}
    t_*=-2.13390\, a\,.
\end{equation}
Now, following~\cite{Emparan:2016ylg} it is possible to calculate the `duration' of the merger, $\Delta_*$, defined as the difference between the horizon's time $t_h$ as it crosses $r=r_*$ (in the direction $\phi =0 $) and the pinch-on instant. An equivalent definition can be given in terms of retarded times, defined as $v=t+\rho$, where 
\begin{equation}
\rho = \int \frac{\sqrt{r^2-a^2}}{r} dr
\end{equation}
is the tortoise coordinate for the metric~\eqref{EllisBmetric}.

In terms of the retarded times, the duration of the merger is given by
\begin{equation}
    \Delta_*:= v_h - v_* = \big(t_h+\rho(r_*)\big) - \big(t_*+\rho(r_*)\big)\,.
\end{equation}
The time $t_h$ is the moment at which the central generator is at $r=r_*$. This is determined by Eq.~\eqref{eq:centralg}. Therefore: 
\begin{equation}
    \Delta_*= \sqrt{r_*^2 - a^2} - t_*\,,
\end{equation}
and using the solutions obtained for $r_*$ and $t_*$ from~\eqref{eq:rdotequalzero}, it follows that $\Delta_* =2.31244\ a$.

The collision of the Ellis-Bronnikov wormhole with a large BH thus seems to be more than two times faster than the merger of two Schwarzschild BHs in the extreme mass ratio limit. In the latter case it was found that $\Delta_*=5.94165\, r_0$ \cite{Emparan:2016ylg}, where $r_0$ denotes the horizon radius of the small black hole.\footnote{In making this comparison, we are assuming that the sizes of the small objects are the same, i.e., $r_0=a$.}

\section{Merger of a large BH with an intra-universe wormhole\label{sec:intra_universe}}

Considering an intra-universe wormhole (i.e., a wormhole with two mouths in the same universe) leads to novel physical effects regarding the study of the event horizon. Indeed, we will show that events which in the setup of Sec.~\ref{sec:inter_universe} were inside the event horizon might now be part of the domain of outer communications of the BH.

\begin{figure}[t!]
\centering
\includegraphics[width=0.85\linewidth]{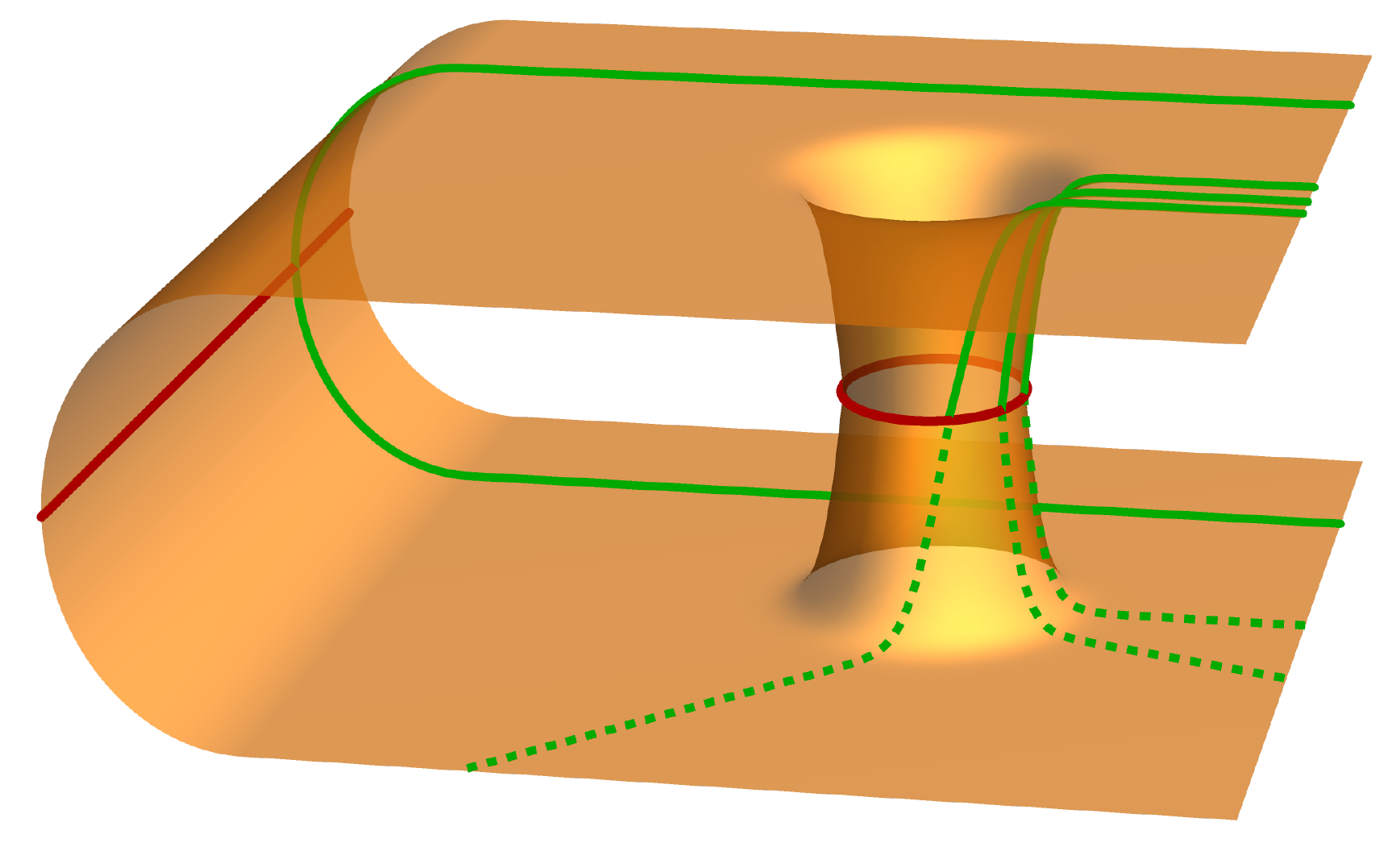}
\caption{Illustration of the intra-universe wormhole. The wormhole connects two distant parts of the same universe through the throat. The red lines represent two surfaces along which we join two halfs of the Ellis-Bronnikov spacetime.}
\label{fig:intraWH}
\end{figure}

Our setup is as follows (see Fig.~\ref{fig:intraWH}). We assume that the axis connecting the two mouths of the wormhole is perpendicular to the planar horizon. Other configurations might be considered, and the precise results should depend on the details of the setup. But qualitatively, the outcome of this investigation is robust, except for the special case in which the two mouths fall through the horizon synchronized.

\begin{figure*}[ht!]
\centering
\includegraphics[width=0.8\linewidth]{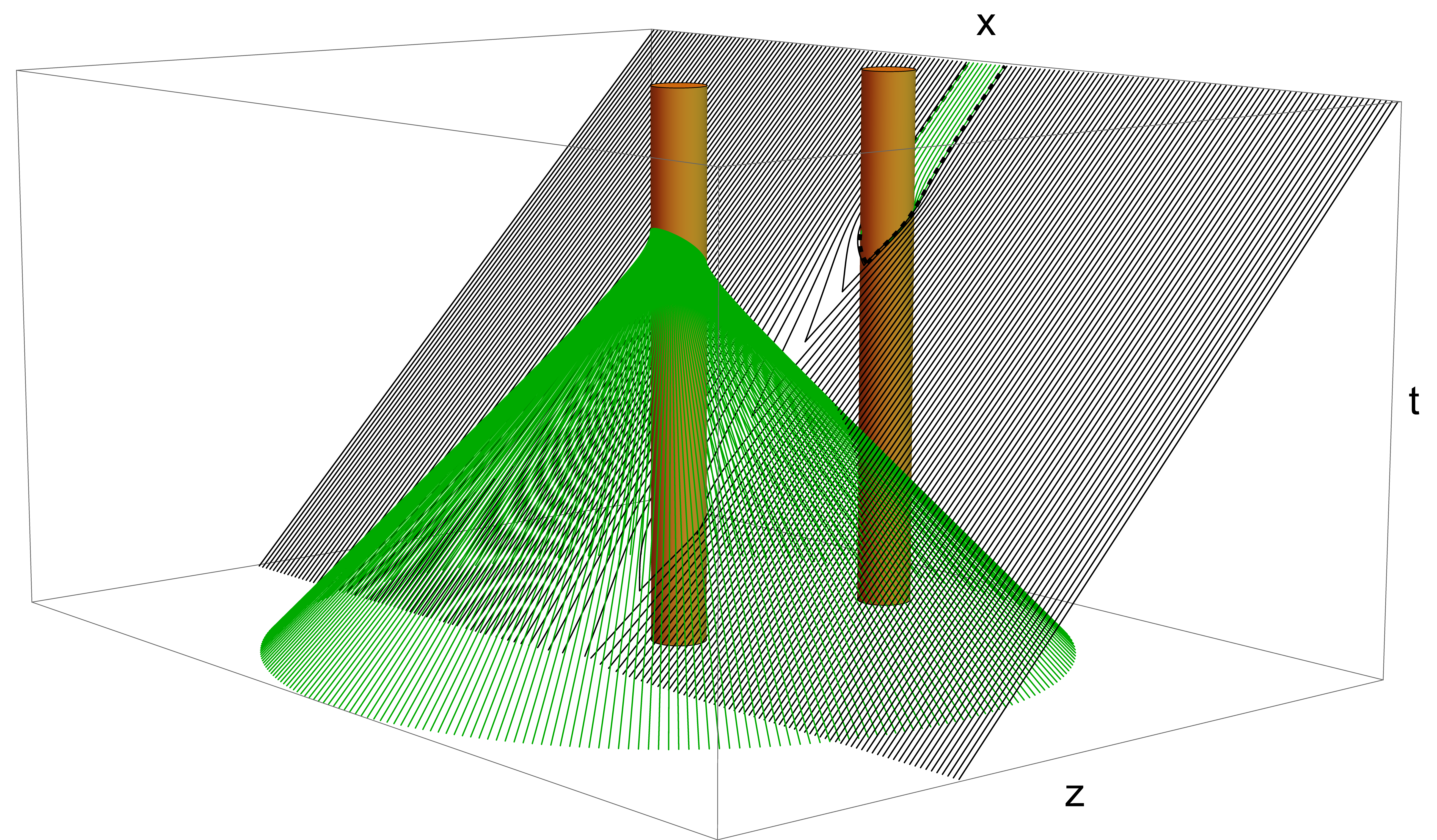}
\caption{Equatorial section of the event horizon as a function of time for an intra-universe wormhole. The distance between the two mouths was set to $d=10\,a$ to produce this plot. The color coding is the same as for Fig.~\ref{fig:ebmtmerger}, except now two infinite cylinders are included, to indicate points that do not belong to the spacetime, as they would be `inside' the wormhole throat. The surfaces of the two cylinders should be thought of as being identified. The green `cone' and the black `plane' intersect along a certain curve ---not computed, nor shown, explicitly--- which is a second caustic line that arises in this model. Even though the continuation of the green and black generators to the past of this second caustic is shown, it must be noted that those sections of the generators are not actually part of the event horizon, and the black hole interior is the region that lies simultaneously above the green `cone' and the black `plane'. The lifetime of the island is the difference between the last instant at which the green `cone' and the black `plane' intersect, and the last instant at which the green `cone' still exists.}
\label{fig:intra3d}
\end{figure*}

With this assumption, the closest wormhole mouth falls into the large black hole before the other one. In the absence of the wormhole no light rays emitted from the black hole region can ever reach asymptotic null infinity. Once we include the in-falling wormhole, some of these light rays can enter the leading mouth of the wormhole and exit from the trailing mouth while it is still outside the BH.
Note that this is possible since we are also assuming that the wormhole has zero length: a signal that enters one mouth \textit{immediately} comes out the other mouth. If the wormhole length is strictly positive, this will cause some delay. And if the length of the wormhole is greater than the direct path connecting the two mouths ---as is imposed by the achronal averaged null energy condition (AANEC)~\cite{Galloway:1999bp, Gao:2000ga, Graham:2007va, Wall:2009wi, Kontou:2020bta}---, then the in-falling wormhole does not open any window to observe the interior of the large black hole, as we demonstrate below.

\begin{figure}[hbt!]
\centering
\includegraphics[width=0.95\linewidth]{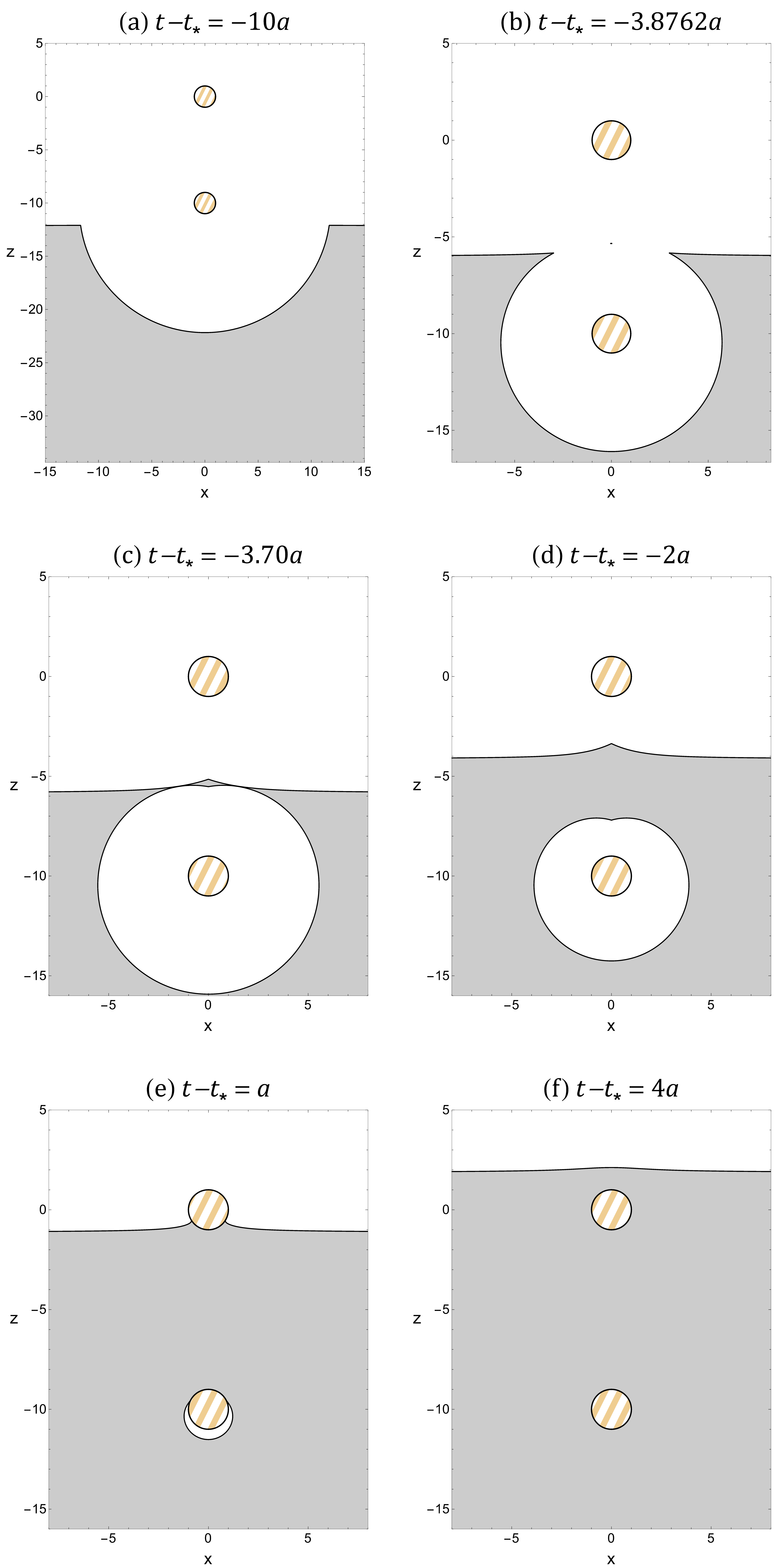}
\caption{Time slices of the event horizon for an infalling intra-universe wormhole. As in Fig.~\ref{fig:intra3d}, the inter-mouth distance is set to $d=10\,a$ and the value of $t_*$ (defined in Sec.~\ref{subsec:intra-B}) is the same as in Sec.~\ref{sec:inter_universe}, namely $t_*=-2.13390\,a$. Among the six panels shown we highlight two. Panel $(b)$ is a still shot of the event horizon at the moment of the formation of the precursor, which is visible as a tiny dot near $(x,z)=(0,-5)$. Panel $(c)$ illustrates the configuration occurring when the precursor merges with the rest of the black hole region, and it corresponds to the birth of the island.}
\label{fig:intraslices}
\end{figure}

The assumption we made, i.e, that the axis connecting the two mouths of the wormhole is
perpendicular to the planar horizon,
also has the advantage of preserving the axial symmetry of the congruence of null generators that form the event horizon. This allows us to obtain all the null geodesics as a function of a single parameter, namely the already mentioned impact parameter $q$. If this were not the case we would need two parameters to describe the whole family of generators.

In order to build a global spacetime with two wormhole mouths in the same universe, we will glue two equal copies of the Ellis-Bronnikov metric along a plane (at the same distance from both mouths), as illustrated in Fig.~\ref{fig:intraWH} with the red straight line. This is effectively assuming that there is a domain wall between the two mouths, and it must have some form of negative energy ---which can be computed--- to prevent the two mouths from falling into each other. Nevertheless, if the mouths are sufficiently far away, the effect of the negative energy domain wall can be considered negligible.

\subsection{Horizon generators}

We shall use the results of Sec.~\ref{sec:inter_universe} to obtain the generators around each of the mouths. The only additional part of the construction that needs to be described concerns the domain wall. Namely, we should specify how we continue the geodesics across the matching plane. 
Assuming the effect of the domain wall to be negligible, the geodesics should simply be continuous and differentiable at that surface.
This is what we implemented in order to obtain Fig.~\ref{fig:intra3d}, where we chose the distance between the two wormhole mouths to be $d=10\,a$, for concreteness.

Differently from Sec.~\ref{sec:inter_universe}, the generators that emerge from the trailing wormhole mouth are the continuation of null geodesics that entered the leading mouth \emph{in the same universe}. Therefore, the green `cone' coexists with the black `plane'. However, contrary to the presentation of Fig.~\ref{fig:ebmtmerger}, we now rotate the green `cone' around the vertical axis by $180^\circ$. This is done in order to reproduce the intuition provided by Fig.~\ref{fig:intraWH}. Specifically, we want the geodesics exiting the trailing mouth (on its right side) to be connected with geodesics entering the leading mouth from its left side.

In this intra-universe wormhole setup there are new caustic lines not aligned with the axis $x=0$. These caustics arise from the intersection of the black `plane' with the green `cone'. This is apparent from Fig.~\ref{fig:intra3d}. Obtaining them is more challenging than the caustics we computed in the previous section, and is not the main purpose of our analysis.
It is important to realize that the interior of the black hole corresponds in Fig.~\ref{fig:intra3d} to the region above \emph{all} the generator congruences (i.e., both the black and the green surfaces).

\subsection{Physical interpretation}
\label{subsec:intra-B}

Let us now describe the whole evolution of the merger process.

Upon taking time slices of Fig.~\ref{fig:intra3d}, at early times (bottom of the figure) we see an almost flat horizon, when the two wormhole mouths are far from the large BH, but with an arc of a circle `cut out' from the BH interior (see also Fig.~\ref{fig:intraslices}). As time progresses, the horizon gets more and more deformed, as shown in Fig.~\ref{fig:intraslices}. Note that, in this intra-universe setup, we define $t_*$ as the last moment still featuring a caustic point along the event horizon. I.e., it is the maximum $t$-coordinate of the caustic line. This definition applies equally well to Sec.~\ref{sec:inter_universe}, but now we are considering two distinct wormhole mouths in the same universe, so we may specify that it corresponds to the saddle point on the congruence of generators closest to the trailing wormhole mouth.\footnote{There would also exist a saddle point on the congruence of generators closest to the leading wormhole mouth, but that is not part of the event horizon. See Fig.~\ref{fig:intra3d}.}
An important difference with respect to Sec.~\ref{sec:inter_universe} is that, contrary to what happens for the inter-universe WH, in this case the black hole horizon bends \emph{away} from the leading WH mouth --- while it bends toward the trailing WH mouth.

At a given time, $t_p$, a \emph{precursor} forms between the two mouths. This precursor is a spatial region that belongs to the BH but is disconnected from the rest of its interior. (See panels $(b)$ and $(c)$ in Fig.~\ref{fig:intraslices}.) The same phenomenon appears in the merger of a neutron star with a large black hole \cite{Emparan:2020uvt}.
The precursor expands until it connects, at a later time, $t_i>t_{p}$, with the large horizon, thus closing off an `island' region.
This island is spatially located behind the large horizon but is actually not part of the black hole interior because one can escape the BH through the WH.
As time further elapses, the island shrinks until it disappears into the wormhole mouth. 
Finally, when the two mouths are both inside the BH, the WH ceases to offer an escaping route and the horizon flattens out again.

The possibility of creating an island is a consequence of our assumptions. In particular, we chose to connect the two mouths with a zero length bridge which effectively provides a shortcut in spacetime.
Imposing the physically-relevant achronal averaged null energy condition amounts to making the length through the wormhole longer than the direct path without crossing the wormhole~\cite{Graham:2007va, Maldacena:2018gjk, Emparan:2021xdy}. In practice, this will move down the green cone in Fig.~\ref{fig:intra3d} in such a way that the novel effects we describe in this section (the precursor and the island) will be limited to a region closer to the leading mouth. If the length of the WH is taken to be too long, these effects will be completely absent.
This statement might need to be revised if a twist along the wormhole is included, as discussed in section~\ref{sec:conclusions}.

We emphasize that the islands we discuss here have no direct relation with the quantum extremal islands~\cite{Engelhardt:2014gca, Penington:2019npb, Almheiri:2019psf, Geng:2021hlu} that have been recently proposed, in the context of the black hole information paradox, to reproduce the Page curve for evaporating black holes~\cite{Almheiri:2019hni, Akers:2019nfi, Penington:2019kki, Almheiri:2019qdq, Almheiri:2020cfm}. Despite some clear differences at the level of the details, the spirit is, nevertheless, the same: there is a region within the black hole that is, in a sense, not part of it.

\vfill

\subsection{Quantitative characterization}

As already described, in this setup the merger between the two objects starts with the formation of a precursor. The time evolution of the whole process can be obtained by taking constant time slices of Fig.~\ref{fig:intra3d}, and this is illustrated in Fig.~\ref{fig:intraslices}.

In this section, we characterize this process by computing the island lifetime, $\Delta_\circ$, 
 which measures the time between the precursor formation and the collapse of the island. 
To be precise, this is a slight overestimate because the island only forms when the precursor fuses with the large horizon. However, that instant is difficult to compute exactly, since the pinch-on does not occur along the $x=0$ axis (see Fig.~\ref{fig:intraslices}, panel $(c)$). Given that the precursor remains disconnected from the rest of the black hole region only for a very short time when compared to the lifetime of the island ---especially for large $d\gg a$---, this is a reasonable approximation.

While the last instant for which the island still exists is $t_f=0$ ---obtained from Eq.~\eqref{eq:centralg} with $r=a$---, the moment of birth of the island, $t_i$, is much harder to determine. We will use, as a proxy, the instant at which the two caustic lines along $x=0$ cross. I.e., we shall take $t_i=t_p$. The characteristic lifetime of the island can then be defined, naturally, as $\Delta_\circ = t_f - t_i$.

Once $d$ is fixed relative to $a$, the lifetime can be computed numerically. We have calculated it for a range of ratios $d/a$ and, as expected, a linear law of the form
\begin{equation}
\label{eq:lifetime1}
\Delta_{\circ} = a+\frac{1}{2} d
\end{equation}
yields a good fit, with a precision better than $2.5\%$ for all inter-mouth distances $d$ greater than $4a$.
This linear fit remains appropriate  even for small inter-mouth distances, $d/2-a\ll a$, although that would mean the two mouths are very close to each other. As a result, the domain wall keeping them apart would have a strong impact on the generators, and so such cases are outside the regime of validity of our analysis.

As mentioned before, we adopted a zero-length wormhole to conduct our calculations. If one considers a wormhole with a finite length $L$, a time corresponding to half this length must be deducted from the island's lifetime, when the intermouth distance is much larger than the size of the throat, so the previous relation must be modified according to
\begin{equation}
\label{eq:lifetime2}
\Delta_\circ \simeq \frac{d-L}{2} \,, \quad \text{for} \;\; d\gg a\,.
\end{equation}

The lifetime of the island ---as we defined it above--- does not employ the retarded time, in contrast with what was done in sec.~\ref{sec:inter_universe} for the duration of the merger. Unfortunately, in this setting a similar definition using retarded time appears to be out of reach.\footnote{The main obstacle to defining a retarded time associated to the birth of the island is a simple topological fact: in such a spacetime, light rays emitted along the axis $x=0$ from that event, corresponding to the intersection of the caustic lines, can never reach infinity. Instead, they would keep eternally cycling through the wormhole. In addition, the spacetime is not spherically symmetric, which also poses a difficulty of technical nature.}

It should be noted that imposing the AANEC amounts to requiring that $L>d$, so that the wormhole is `long'~\cite{Graham:2007va, Maldacena:2018gjk, Emparan:2021xdy}. Eq.~\eqref{eq:lifetime2} then implies that the lifetime would be negative, meaning that the island is necessarily absent under these conditions.

\section{Conclusions and Discussion \label{sec:conclusions}}

In this work, we have studied the merging process between a large black hole and two different wormholes in the extreme mass ratio limit; namely, in Sec.~\ref{sec:inter_universe} a single-mouth wormhole, and in 
Sec.~\ref{sec:intra_universe} a two-mouth wormhole. The whole analysis was performed using the Ellis-Bronnikov metric and employing ray-tracing techniques.
Both setups present novel features together with some already well-described phenomena occurring in the black hole-black hole~\cite{Emparan:2016ylg} and the black hole-neutron star fusions~\cite{Emparan:2020uvt}. 
We have shown that basically all of the features observed in previous studies of (non-rotating) mergers in the EMR limit are ---in some cases surprisingly--- present also here: most notably, the fusion of initially disconnected horizons in Sec.~\ref{sec:inter_universe}, and the appearance of precursors in Sec.~\ref{sec:intra_universe}.

The inter-universe wormhole case resembles the merging of two black hole horizons, but, in this case, the finite horizon with $S^2$ topology lives almost entirely on the side of the universe beyond the wormhole throat, and it is the \emph{exterior} of that surface that forms part of the black hole region. 
In this collision, as in a binary black hole merger, two initially disconnected event horizons fuse to become a single connected surface at late times. The point at which the two horizons first touch occurs very near the throat of the wormhole, at least for the Ellis-Bronnikov geometry we adopted. Let us stress again that this background geometry has no horizon; therefore, one might amusingly call such a situation ``the fusion of two horizons in a spacetime without horizons''.

Building on the results of Sec.~\ref{sec:inter_universe}, we have performed the calculations also for the collision with an intra-universe wormhole, in Sec.~\ref{sec:intra_universe}. This setup shows the possibility of creating an `island' --- a region behind the infinitely large horizon which, nevertheless, does not belong to the black hole interior. This occurs because signals emitted from within can escape the black hole by falling into the leading mouth of the wormhole and exiting from the trailing mouth, which is still traveling through the domain of outer communications. The life-time of such islands depends essentially linearly on the distance between the two mouths of the wormhole and on the wormhole length.

Many of the features borne out of the calculations performed in Sec.~\ref{sec:intra_universe} can be inferred from a simpler model of a wormhole, which is obtained by identifying two slits in a flat spacetime.\footnote{We thank Roberto Emparan for having pointed out to us the similarities with this model.}
That simpler model immediately makes it apparent that the island lifetime scales like half the distance between the two mouths. The more involved model we have used allowed us to compute corrections to that linear law stemming from the curvature of the spacetime. As expected, in the limit of large separations the two expressions agree. What is somewhat surprising is that even at relatively small inter-distances compared to the throat size, the departure from linearity continues to be negligible.

A notable difference between the results of sections~\ref{sec:inter_universe} and~\ref{sec:intra_universe}, apart from the appearance or not of an island, concerns the bending of the large horizon. While the geometry of the wormhole dictates that the large horizon bends towards the wormhole in Sec.~\ref{sec:inter_universe}, once a shortcut through spacetime is created by connecting two mouths within the same universe, as in Sec.~\ref{sec:intra_universe} , it follows that the large horizon recedes away from the leading mouth of the wormhole, as expected.

It is interesting and instructive to compare our findings with the original study by Frolov and Novikov~\cite{Frolov:1993jq}, which proposed a wormhole falling into a black hole as a gedanken experiment to access the interior of the BH.
As in our case, those authors considered a traversable wormhole in the limit of vanishingly small mass for the mouths and zero length for the handle between the two mouths.
However, in contrast with our analysis, they investigated the propagation of null geodesics in the Schwarzschild spacetime (instead of the EB background), and explicit results were obtained by assuming the wormhole mouths to be point-like. Therefore, Ref.~\cite{Frolov:1993jq} also explored the extreme mass ratio regime, although in a different guise: in~\cite{Frolov:1993jq} it is the large BH whose size is kept finite while the WH size is taken to zero; in the present paper we chose to keep the size of the WH finite and let the BH horizon become infinitely large.

Given that both studies analyze the same gravitational system in similar regimes, one should expect agreement between the main results. This is indeed the case.
Ref.~\cite{Frolov:1993jq} showed how the absorption of a wormhole generates a temporary shrinking of the black hole's horizon while leaving the outside gravitational field largely unchanged. This is the same feature we observed in Sec.~\ref{sec:intra_universe}. However, the presence of a precursor and the related island went unnoticed in that case.
Nevertheless, a careful reading of~\cite{Frolov:1993jq} demonstrates that their analysis also supports our result about the existence of an island. The crucial point is that Ref.~\cite{Frolov:1993jq} adopted Eddington-Finkelstein coordinates, where the time-like coordinate is replaced by a light-like coordinate. The upshot is that the exterior of the black hole in those coordinates is connected. But if one were to revert back to a time-like coordinate, slices of constant time would be tilted and then it becomes clear that, for a certain time interval, there is a region contained within the outermost horizon which is, nonetheless, part of the exterior of the black hole. This is the island.

The intriguing conclusion of~\cite{Frolov:1993jq} was that it is possible to extract information from the interior of a black hole if only one allows a tiny violation of the weak energy condition, since drastic modifications to the event horizon can be produced by a WH as small as desired. The analysis we have presented allows a full description of the changes suffered by the event horizon on the scales of the wormhole that causes its deformation.
Our results support the expectation that, once physically reasonable energy conditions are imposed (i.e., the achronal averaged null energy condition, which forbids the existence of short wormholes), then our ability to recover information classically from within a black hole by using WHs is lost. In any case, we hope this study can serve as a useful testbed for ideas concerning the information paradox.

Let us finally discuss possible future directions. We have made a number of choices in the modelling of the intra-universe wormhole, so a natural extension would be to consider other configurations. Quantitatively, the results will depend on the details of the setup. We have already considered changes to the distance between the two mouths (relative to the size of the wormhole mouth), and the length of the wormhole. Other parameters of the model that might play a role are the twist along the wormhole, and the orientation of the axis connecting the two mouths compared to the planar horizon. For example, there could be a chance to create a short-lived island without any violation of the AANEC if the wormhole twists. Although this seems unlikely, it is still an open possibility that requires a dedicated analysis.

One may also consider more drastic modifications to the setup, for instance by delaying the plunge of the trailing wormhole mouth or even forcing it to remain fixed outside of the large BH, while the leading mouth falls in. At first sight, the latter variant seems to entirely destroy the horizon, since one could always escape the black hole by choosing a route through the wormhole to the exterior. However, this conclusion is also an artifact of the assumed extreme mass ratio regime, namely that the black hole was taken to be infinitely large, which implies that its own strong gravitational field is localized infinitely far away from the horizon.

The Ellis-Bronnikov solution we adopted as a background in this paper features a single parameter, $a$, that simultaneously controls the size of the wormhole and the curvature of the spacetime. However, a generic wormhole solution should allow these two parameters to vary independently. For example, Ref.~\cite{Bakopoulos:2021liw} derives wormholes which are massive and for which one can freely adjust the size of the throat. See also~\cite{Bronnikov:2021liv} for other such spacetimes. 
Another possibility is to consider ``portal'' wormholes \cite{Feng:2021pfz}, which can be designed in such a way that the curvature vanishes everywhere outside a negative mass cosmic string that supports the wormhole. The flatness, combined with the non-trivial topology of such spacetimes, suggests that such a setup might provide a particularly simple means to reveal many of the essential features we have studied, in the same spirit as the two-slit model previously mentioned. 

 A related topic is the study of the marginally outer trapped surfaces (both open and closed, possibly self-intersecting), which can also be considered in the extreme mass ratio case. Following~\cite{Booth:2020qhb}, the whole analysis that we have conducted here can be supplemented with the study of the properties of such  trapped surfaces. Given that the event horizon must enclose all marginally outer trapped surfaces, it might be interesting to explore the connection between the latter and the constant time slices we have constructed featuring islands.

Finally, it would be worthwhile to improve the construction of the global solution with two wormhole mouths in the same universe. We achieved that in Sec.~V by gluing two copies of the same EB geometry along a domain wall, arguing that if the mouths are sufficiently far apart, the negative energy localized on that surface is negligibly small. Still it would be desirable to employ an exact solution without requiring any additional matter sources. Presently such intra-universe wormhole solutions are not known exactly and it would be very interesting to derive such geometries, possibly making use of a high degree of symmetry in order to make the problem more tractable.

\medskip

\begin{acknowledgments}
We thank Roberto Emparan, Robie Hennigar and Orlando Panella for useful discussions. AMF is supported by ERC Advanced Grant GravBHs-692951, MICINN grant PID2019-105614GB-C22, AGAUR grant 2017-SGR 754, and State Research Agency of MICINN through the “Unit of Excellence Mar\'ia de Maeztu 2020-2023” award to the Institute of Cosmos Sciences (CEX2019-000918-M). JVR acknowledges support from FCT under projects 2022.08368.PTDC, CERN/FIS-PAR/0023/2019, UIDB/00099/2020 and UIDP/00099/2020.
\end{acknowledgments}

\appendix

\section{Derivation of caustic lines}
\label{app:AppendixA}

In this appendix, we provide explicit expressions for the integrals in Eqs.~\eqref{eq:integ_t} and~\eqref{eq:integ_phi}, and use them to compute the caustic line occurring along $\phi=\pi$.

Before presenting the formulae, it is useful to stress that the background geometry only has a single length scale, namely the wormhole throat radius, $a$. Therefore, all dimensionful quantities can be written entirely in terms of the ratios $\xi=|q|/a$ and $\nu=r/a$. We also introduce the dimensionless time coordinate, $\eta=t/a$.

Upon fixing the integration constants $\alpha_q=\beta_q=0$, as discussed in section~\ref{subsec:inter_universe_generators}, and using results from Ref.~\cite{Pina:2022dye}, it follows that

\begin{widetext}

\begin{figure}
\centering
\begin{minipage}{.48\textwidth}
  \centering
  \includegraphics[width=.9\linewidth]{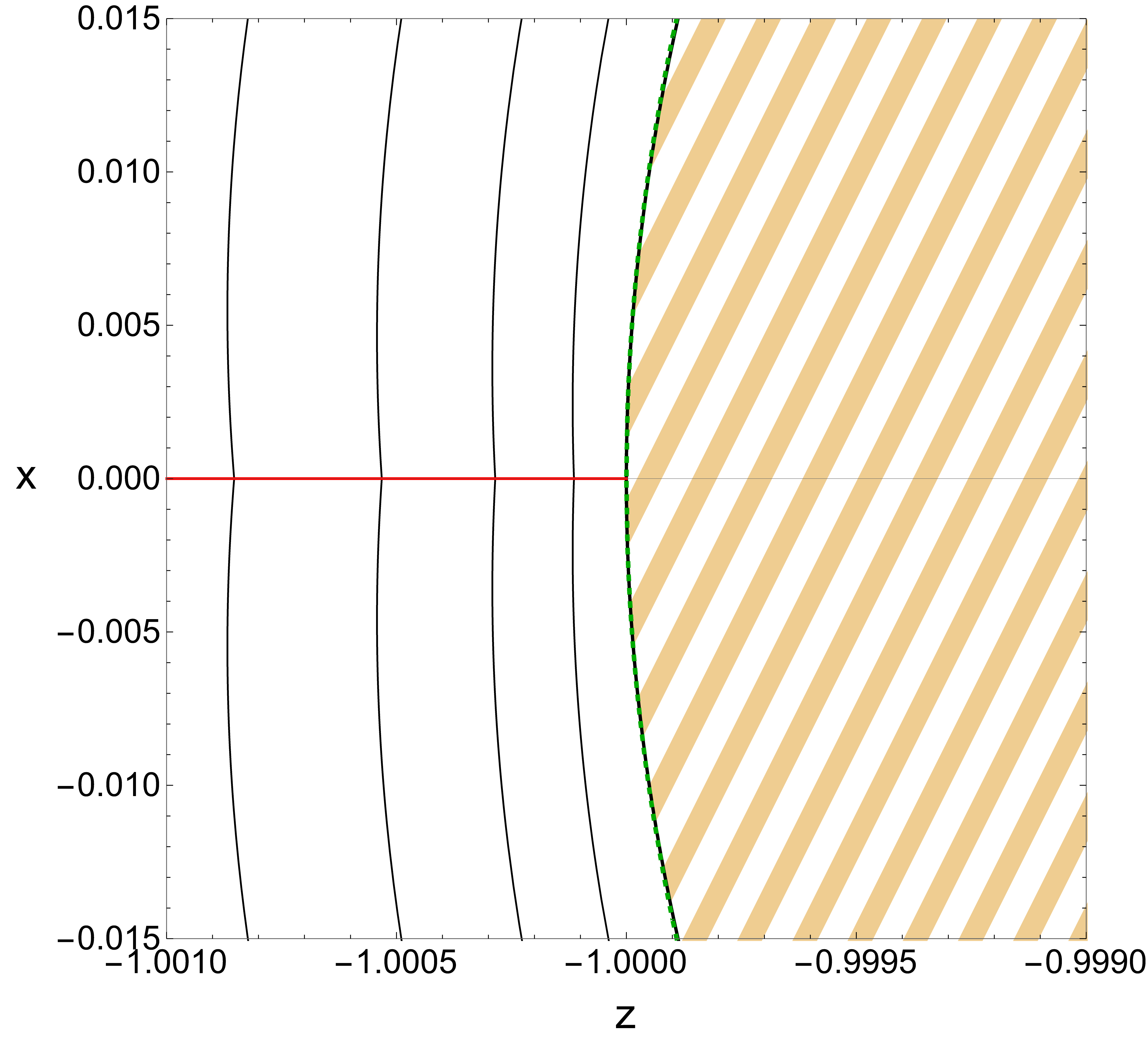}
  \caption{Close up of Fig.~\ref{fig:projection_bothsides}, zooming in on the point at which the generators with the critical impact parameter $q_c$ (shown in dashed green --- not to be confused with the throat of the wormhole) reach the throat radius. All other generators shown meet the $x$-axis in an oblique fashion, but the $q=q_c$ generators reach it perpendicularly. The only other generators for which this happens are obtained for $q=q_*$ but they fall outside this zoomed-in region. The horizontal red line is the caustic.}
  \label{fig:2dzoom}
\end{minipage}
\hfill
\begin{minipage}{.48\textwidth}
  \centering
  \includegraphics[width=.9\linewidth]{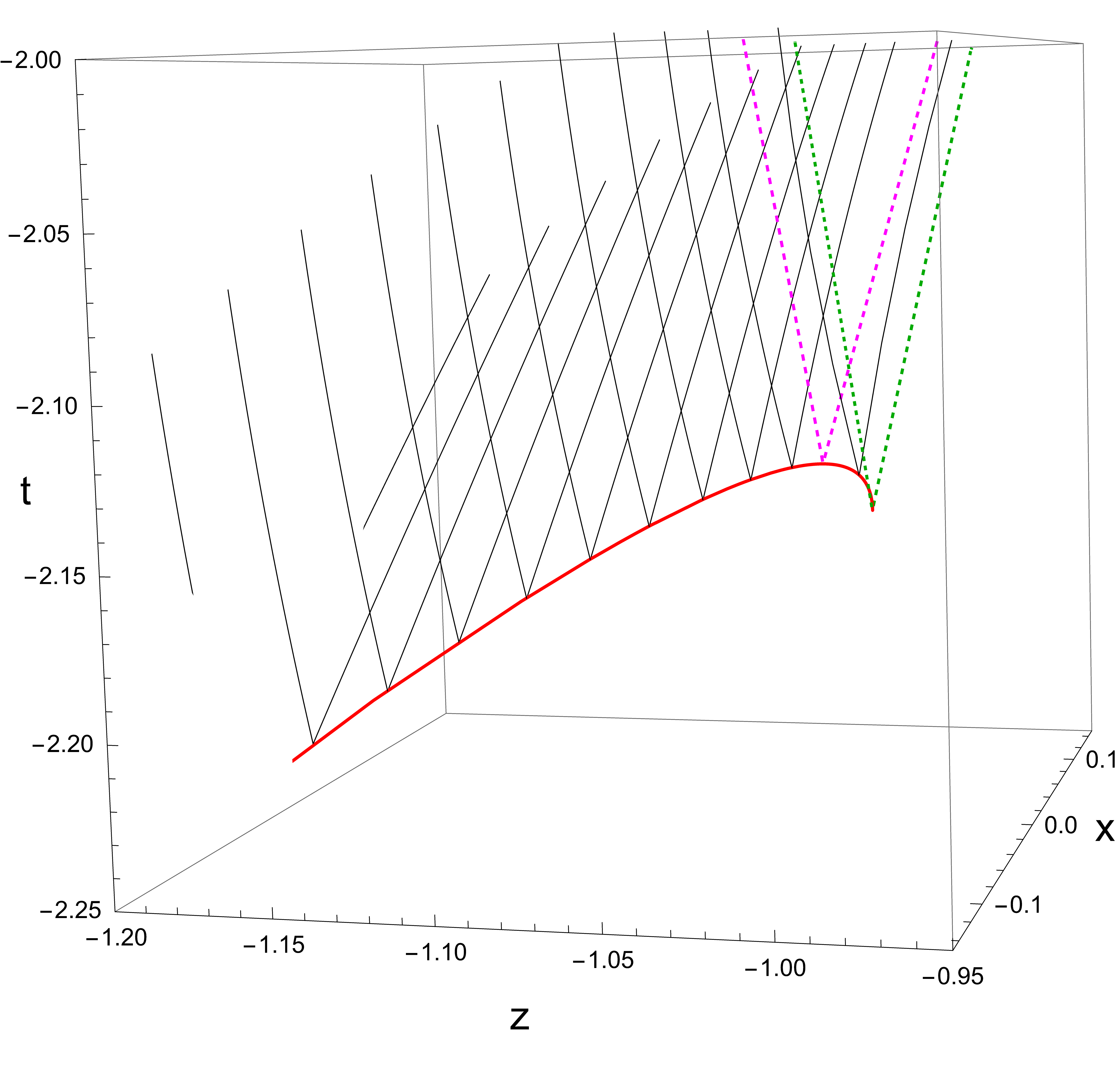}
  \caption{Close up of the left panel of Fig.~\ref{fig:ebmtmerger}, zooming in on the point $(t,r,\phi)=(t_*,r_*,\pi)$, corresponding to a saddle on the event horizon, which occurs very close to the location of the throat ($r=a$). The event horizon is a non-differentiable surface along the caustic line, shown in red. The magenta dashed lines identify the $q=q_*$ generators, while the green dashed lines correspond to the $q=q_c$ generators.}
  \label{fig:zoomsaddle}
\end{minipage}
\end{figure}

\begin{eqnarray}
\label{eq:etavsnu}
    \eta_q(\nu)&=&\frac{t_q}{a}(\nu)= \sqrt{\frac{(\nu^2-\xi^2)(\nu+1)}{\nu-1}} + \sqrt{\xi} \, F\left(\gamma(\nu)\Big|\beta\right) - 2\sqrt{\xi} \, E\left(\gamma(\nu)\Big|\beta\right)-c_q\,,\\
    \phi_q(\nu)&=& \text{sgn}(q) \sqrt{\xi} \left[ F\left(\alpha\Big|\beta\right) - F\left(\gamma(\nu)\Big|\beta\right) \right] \,,
\end{eqnarray}
where the additive constant $c_{q}$ is equal to
\begin{equation}
    c_q=1+\sqrt{\xi} \, F\Big(\alpha \Big|\beta\Big) -2\sqrt{\xi} \, E\Big(\alpha\Big|\beta\Big)\,,
\end{equation}
and
\begin{equation}
    \alpha=\arcsin\left(\sqrt{\frac{2}{\xi+1}}\right)\,, \qquad
    \beta=\frac{(1+\xi)^2}{4\xi} \,, \qquad
    \gamma(\nu)=\arcsin\left(\sqrt{\frac{2(\nu-\xi)}{(\xi+1)(\nu-1)}}\right) \,.
\end{equation}
The functions
\begin{eqnarray}
\label{eq:ellipticF}
    F(x|y)&=&\int_0^x \frac{d\theta}{\sqrt{1-y \sin^2\theta}}\,,\\
    E(x|y)&=&\int_0^x\sqrt{1-y \sin^2\theta}\,  d\theta\,,
\end{eqnarray}
are elliptic integrals of the first and second kind, respectively.

The explicit results given above for the $t$ and $\phi$ coordinates of the generators clearly break down when $r\to a$, but are valid otherwise.

Setting $\phi_q(r)=\pi$ one can determine the $r$-coordinate of the caustic line as a function of the impact parameter $q$. The result, again following~\cite{Pina:2022dye}, is written in terms of the sine of the Jacobi amplitude, $\text{sn}(x|y)=\sin\left(\text{am}(x|y)\right)$, where
\begin{equation}
    \text{am}\big(F(x|y)\big|y\big)=x\,.
\end{equation}
The final expression can be written as
\begin{equation}
\label{eq:nuvsxi}
    \nu_c(\xi)=\frac{r_c}{a}(\xi)=\frac{2\xi-\left(1+\xi\right) \text{sn}^2\left(F\left(\alpha\big|\beta\right)-\frac{\text{sgn}(q)\pi}{\sqrt{\xi}}\;\Big|\;\beta\right)}{2-(1+\xi) \,\text{sn}^2\left(F\left(\alpha\big|\beta\right)-\frac{\text{sgn}(q)\pi}{\sqrt{\xi}}\;\Big|\;\beta\right)}\,.
\end{equation}
Accordingly, the $t$-coordinate of the caustic line, as a function of $q$, is obtained by plugging the above result in Eq.~\eqref{eq:etavsnu}, i.e., $\eta_c(\xi)=\eta_q(\nu_c(\xi))$, or, restoring dimensions,  $t_c(q)=t_q(r_c(q))$\,. 

The value reported in Eq.~\eqref{eq:qstar2} is obtained by equating expression~\eqref{eq:nuvsxi} to $\xi$.

Figs.~\ref{fig:2dzoom} and~\ref{fig:zoomsaddle} are zoomed-in versions of Figs.~\ref{fig:projection_bothsides} and~\ref{fig:ebmtmerger} in the main text, obtained by employing the results of this Appendix. They display the behavior of the caustic line close to the wormhole throat and emphasize that the $q = q_*$ generators are distinct from the $q = q_c$ generators.

\end{widetext}

\bibliography{ref}

\end{document}